\documentclass[aps,prb,showpacs,twocolumn,superscriptaddress]{revtex4}

\usepackage{graphicx}
\usepackage{amsmath}
\usepackage{natbib}
\usepackage{bm}
\usepackage[dvips]{color}

\begin{document}

\title{Hyperfine interactions at lanthanide impurities in Fe}
\date{\today}

\author{D.~Torumba}
\affiliation{Instituut voor Kern- en Stralingsfysica, Katholieke
Universiteit Leuven, Celestijnenlaan 200 D, B-3001 Leuven,
Belgium}
\author{S.~Cottenier}
\email[Electronic address: ]{Stefaan.Cottenier@fys.kuleuven.ac.be}
\affiliation{Instituut voor Kern- en Stralingsfysica, Katholieke Universiteit
Leuven, Celestijnenlaan 200 D, B-3001 Leuven, Belgium}
\author{V.~Vanhoof}
\affiliation{Instituut voor Kern- en Stralingsfysica, Katholieke
Universiteit Leuven, Celestijnenlaan 200 D, B-3001 Leuven,
Belgium}
\author{M.~Rots} \affiliation{Instituut voor Kern- en
Stralingsfysica, Katholieke Universiteit Leuven, Celestijnenlaan
200 D, B-3001 Leuven, Belgium}

\begin{abstract}
The magnetic hyperfine field and electric-field gradient at isolated lanthanide impurities
in an Fe host lattice are calculated from first principles, allowing for the first time a
qualitative and quantitative understanding of an experimental data set collected over the
past 40 years. It is demonstrated that the common Local Density Approximation leads to
quantitatively and qualitatively wrong results, while the LDA+U method performs much
better. In order to avoid pitfalls inherent to the LDA+U method, a careful strategy had to
be used, which will be described in detail. The lanthanide 4f spin moment is found to
couple antiferromagnetically to the magnetization of the Fe lattice, in agreement with the
model of Campbell and Brooks. There is strong evidence for a delocalization-localization
transition that is shifted from Ce to at least Pr and maybe further up to Sm. This shift is
interpreted in terms of the effective pressure felt by lanthanides in Fe. Implications for
resolving ambiguities in the determination of delocalization in pure lanthanide metals
under pressure are discussed. For the localized lanthanides, Yb is shown to be divalent in
this host lattice, while all others are trivalent (including Eu, the case of Tm is
undecided). The completely filled and well-bound 5p shell of the lanthanides is shown to
have a major and unexpected influence on the dipolar hyperfine field and on the
electric-field gradient, a feature that can be explained by their $1/r^3$ dependence. An
extrapolation to actinides suggests that the same is true for the actinide 6p shell. The
case of free lanthanide atoms is discussed as well.
\end{abstract}
 \maketitle

\section{Introduction}
\label{sec-intro} A prototype problem in the field of nuclear condensed matter physics is
to determine and understand the magnetic hyperfine field (HFF) at any of the elements of
the periodic table, incorporated as a substitutional impurity in a simple ferromagnetic
host such as bcc Fe. Understanding hyperfine fields forms a critical test for our
understanding of condensed matter. Moreover, they provide a convenient tool for nuclear
physicists to determine nuclear magnetic moments: two features that explain the decades of
experimental\cite{Rao1985} and theoretical\cite{Akai1990,Haas2003} efforts that have been
devoted to this problem. Today, the hyperfine fields of all elements as substitutional
impurities in bcc Fe are well-understood up to about Z=55\, \cite{Akai1984,Akai1985-sp,
Akai1985-d,Korhonen2000,Cottenier2000}. For the heavier 5d impurities, sizeable deviations
between theory and experiment remain\cite{Ebert1990}. The hyperfine fields of very light
impurities at interstitial sites in Fe have been calculated as
well\cite{Akai1987-int,Takeda1993}. Lanthanide impurities in Fe are much less understood,
both experimentally and theoretically. As far as experiment is concerned, it is hard to
obtain \emph{reliable} values for the lanthanide hyperfine fields (and also for their
electric-field gradients, see below). This problem is illustrated by the rather desperate
conclusion of L.~Niesen in a still useful review\cite{Niesen1976} back in 1976:
``\emph{(...) if we cannot perform experiments that yield unambiguous results, we should
better do no experiments at all.}" On the theoretical side, no \emph{ab initio} studies
have been performed yet for lanthanide impurities in Fe (an approach using a model
Hamiltonian is developed in Ref.~\onlinecite{Oliveira2002}). The reason for this lies in a
known failure of the widely used Local Density Approximation (LDA) within Density
Functional Theory (DFT): LDA is not suitable to describe strong electron
correlations\cite{Richter1998,Divis2000,Petit2001,Jalali2002}. As a result, the strongly
correlated and mainly \emph{localized} 4f states in lanthanides are rendered
\emph{itinerant} by LDA. It can therefore be anticipated that for a lanthanide impurity in
bcc Fe, LDA is incapable to describe correctly the interaction between these localized and
strongly correlated 4f electrons and the itinerant 3d states of the host material. This
could be overcome by treating the 4f states as core states, an approach that has been used
in the past with some success for lanthanides\cite{Richter1998,Divis2000,Jalali2002}. In
this way, however, one forces the f-electrons to behave exactly as in free atoms, which is
not entirely correct. An efficient and popular way to improve on the LDA failure without
resorting to fully atomic 4f behavior, is to use the LDA+U
method~\cite{Anisimov1991,Czyzyk1994,Anisimov1993,Petukhov2003,Anisimov1997}. In LDA+U, the
correlation absent in LDA is reintroduced by an on-site Coulomb repulsion parameter U, to
which an \emph{a priori} value has to be assigned. The LDA+U method has been used in the
recent past with considerable success (recent examples are Refs.~\onlinecite{Antonov2002,
Laskowski2003, Petukhov2003, Boukhvalov2003, Shick2003} and many others), but it is not yet
clear where the boundaries of its range of applicability are. In this work, we will examine
how well LDA+U performs on a delicate quantity as the HFF.

An extra feature for lanthanide impurities in Fe that is absent
for lighter impurities, is the presence of a large electric-field
gradient (EFG) at the lanthanide nucleus. At a site with cubic
point symmetry -- such as a substitutional site in bcc Fe -- the
EFG tensor must be necessarily zero. Due to the strong spin-orbit
coupling for these heavy impurities, however, the crystalline
cubic point symmetry at the lanthanide site is lowered to a
tetragonal one. This allows the existence of a large EFG at the
nucleus of the impurity. The same happens for 5d impurities in
Fe\cite{Seewald1997,Seewald1999,Seewald2002}, but there the EFG is
2 orders of magnitude smaller. We have calculated and analyzed
this EFG for lanthanides, and compare it with the sparse
experimental data.

The goals of this work can be summarized as follows. On the physical side, we want to
obtain better \emph{quantitative} and \emph{qualitative} insight in magnetic hyperfine
fields and electric-field gradients of lanthanide impurities in Fe. This should allow us to
asses better the reliability of the existing experiments, and to derive the underlying
physical mechanism. On the technical side, we want to examine whether the range of
applicability of the LDA+U method can be extended to problems as delicate and sensitive as
magnetic and electric hyperfine interactions of heavy impurities in a transition metal
host. It will be shown that in the course of this analysis unexpected new results and
questions show up, such as the influence of the lanthanide 5p electrons
(Sec.~\ref{sec-ldau} and \ref{sec-efgLDAU}) and the position of the
delocalization-localization transition in this system (Sec.~\ref{sec-ldau} and
\ref{sec-pressure}).

\section{Computational details}
\label{sec-compdetails}

All our calculations were performed within Density Functional
Theory\cite{Hohenberg1964,Sham1965,DFT-LAPW2002}, using the
Augmented Plane Waves + local orbitals (APW+lo)
method\cite{Sjostedt2000,Madsen2001,DFT-LAPW2002} as implemented
in the WIEN2k package\cite{wien2k} to solve the
scalar-relativistic Kohn-Sham equations. In the APW+lo method, the
wave functions are expanded in spherical harmonics inside
nonoverlapping atomic spheres of radius $R_{\mathrm{MT}}$, and in
plane waves in the remaining space of the unit cell (=the
interstitial region). For the Fe atoms a $R_{\mathrm{MT}}$ value
of 2.20 a.u.\ was chosen, while for the lanthanide impurity we
used $R_{\mathrm{MT}}=$2.45 a.u. The maximum $\ell$ for the
expansion of the wave function in spherical harmonics inside the
spheres was taken to be $\ell_{\mathrm{max}}=10$. The plane wave
expansion of the wave function in the interstitial region was made
up to
$K_{\mathrm{max}}=7.5/R_{\mathrm{MT}}^{\mathrm{min}}=3.41$~a.u.$^{-1}$,
and the charge density was Fourier expanded up to
$G_{\mathrm{max}}=16 \sqrt{Ry}$.

The lattice constant of Fe was fixed at the experimental value of 2.87~\AA. In order to
reproduce the situation of an isolated impurity in bulk Fe, we used the supercell approach
with a $2\times2\times2$ supercell where one iron atom was replaced by a lanthanide atom.
The neighboring Fe atoms will be displaced by the presence of this impurity, as was
documented before for lighter impurities in Fe\cite{Korhonen2000,Cottenier2000,Haas2003}.
We took this effect into account in an average way by relaxing the nearest neighbors for Eu
as an impurity (which is in the middle of the lanthanide series), and kept the same
relaxation fixed for all other lanthanides. The Eu-Fe distance was 2.60~\AA\, which is an
increase of 0.11~\AA\, with respect to the Fe-Fe distance and which is almost identical to
the distance between 5p impurities and their Fe neighbors\cite{Cottenier2000}. It was
tested for another lanthanide (Er) that there was only a marginal difference of less than
1~T between the HFF obtained with the Eu-Fe distance and the correct Er-Fe distance
(2.58~\AA). A test for an extended supercell of 32 atoms was also performed. We relaxed the
first four nearest neighbors. The Eu-Fe distance hardly changed (2.63~\AA) and the Fermi
contribution to the hyperfine field changed with 5~T. For the sampling of the Brillouin
zone (BZ) of the $2\times2\times2$ supercell we took 75 special k-points in the irreducible
part of the BZ, which corresponds to a $10\times10\times10$ mesh.

As exchange-correlation functional, the Local Density Approximation (LDA)\cite{Perdew1992}
was used.  Spin-orbit (SO) coupling was taken into account in all the calculations by a
second-variational step scheme\cite{Koelling1977}, using a cut-off energy
$E_{\mathrm{cut}}^{\mathrm{SO}}=3.0$~Ry. Relativistic Local Orbitals (RLO) for the
lanthanide 5p states were added to the basis set, because it is known that for actinides
this allows to reduce the basis set size needed for the second variational
step\cite{Kunes2001} (=lower $E_{\mathrm{cut}}^{\mathrm{SO}}$). Limitations in the
implementation prevent to obtain correct EFG's and dipolar HFF's when RLO's are used.
Therefore, whenever such information was needed, the calculations were repeated without
RLO's. This never had a large influence on the obtained values, however. For the LDA+U
method, the `Around the Mean Field' (AMF) scheme of Czy\.{z}yk and
Sawatzky\cite{Czyzyk1994} was used. The choice of the U and J parameters is discussed in
detail in Sec.~\ref{sec-ldau}.

Free atoms were simulated by a supercell containing only one
lanthanide and vacuum otherwise, leading to a separation of
9.4~\AA\ between two `neighboring' lanthanides. For free ion
calculations, this cell was charged. All the other parameters were
chosen exactly the same as in the calculations for lanthanides in
Fe.

\section{Magnetic hyperfine fields}
\subsection{Experimental data set}

\label{sec-expdata} 
\begin{figure}
 \begin{center}
  \includegraphics[width=8cm,angle=0]{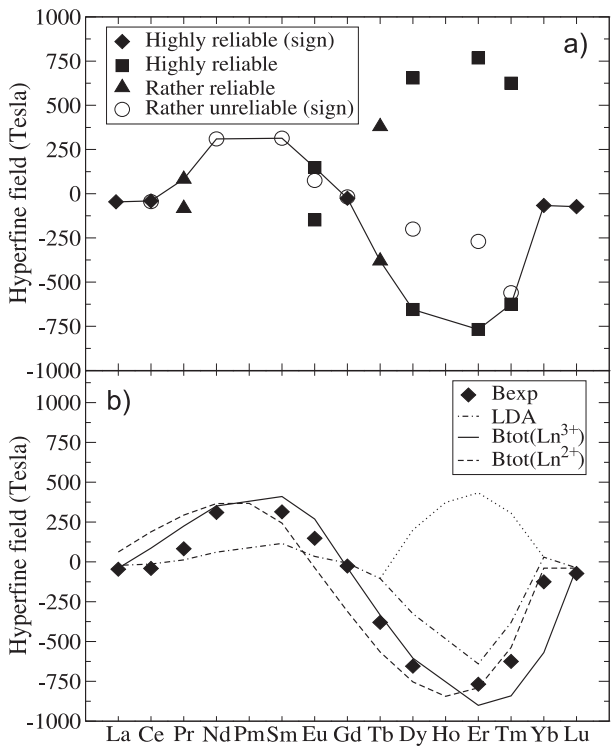}
  \caption{a) Experimental data set for the magnetic hyperfine fields of lanthanides in Fe. If the sign
  of the HFF is not measured, the data point is plotted both at positive and negative values. Distinction
  is made between highly reliable data for which the sign is measured (diamond), highly reliable data without
  sign measurement (square), less reliable data without sign (triangle) and data that are rather unreliable for
  the magnitude of the HFF but reliable for the sign (circle). For references and values, see text. The line connects
  the most likely values for all lanthanides. If multiple measurements with the same reliability were available,
  only one of them is given. More data can be found in the compilation of Rao\cite{Rao1985}. b)~Comparison between
  experiment and several types of calculations for the magnetic hyperfine field of lanthanides in Fe. Diamonds:
  most probable experimental data points (this is the full line from Fig.~\ref{fig1}-a). Dotted-dashed
  line: LDA results for the antiferromagnetic orientation (Campbell-Brooks orientation). Dotted
  line: LDA results for the ferromagnetic orientation when this
  one has the lowest energy. Full line: LDA+U value for trivalent
  lanthanides. Dashed line: LDA+U value for divalent lanthanides. \label{fig1}}
 \end{center}
\end{figure}

Let us first have a look at the experimental
data set for the HFF (Fig~\ref{fig1}-a).
Only in 4 cases the magnitude of the HFF \emph{and} its sign are known with high
reliability (the sign of the HFF indicates whether the field is parallel (+) or
antiparallel (-) with respect to the magnetization of the Fe host lattice). These cases are
La (-47(1)~T)\cite{Goto2003}, Ce (-41(2)~T)\cite{VanRijswijk1983} and Lu
(-73.12(36)~T)\cite{Herzog1985} for which Nuclear Magnetic Resonance on Oriented Nuclei
(NMR/ON) has been performed, and Yb (-125(8)~T)\cite{Devare1978} on which Time-Dependent
Perturbed Angular Correlation spectroscopy (TDPAC) has been applied. The latter technique
has also been used for Gd\cite{Klepper1968}, albeit on a recoil-implanted sample which is
not necessarily clean. The value of -26(8)~T obtained in this way agrees well with an in
principle reliable M\"{o}ssbauer measurement of -37~T, which is unfortunately not very well
documented\cite{Russel1970}. Three time-integrated Perturbed Angular Correlation (IPAC)
measurements are available for Gd as well -- IPAC is a method that is rather unreliable,
and can merely be used to determine the sign and an order of magnitude. They yield
-20(5)~T\cite{Boehm1966}, -18(9)~T\cite{Grodzins1966} and -7~T\cite{Brenn1968}. A HFF of
-30(10)~T can therefore be assigned to Gd in Fe in a reliable way. In 4 other cases the
magnitude of the HFF but not its sign has been measured with an accurate method as
M\"{o}ssbauer Spectroscopy (MS): Eu (148.2(9)~T)\cite{Cohen1974,Niesen1978}, Dy
(610(7)~T)\cite{Wit1978}, Er (768(13)~T)\cite{Niesen1977} and Tm (671~T)\cite{Niesen1976}.
For Pr\cite{Reid1969} and Tb\cite{Niesen1971}, the magnitude but not the sign has been
measured with Low Temperature Nuclear Orientation (LTNO). This non-resonant technique
provides data that are less accurate than the previous ones, although they still are
reasonably reliable. Finally, in the case of Ce\cite{Kugel1971}, Nd\cite{Boehm1966},
Sm\cite{Kugel1976}, Eu\cite{Boehm1966}, Gd\cite{Grodzins1966}, Dy\cite{Grodzins1966},
Er\cite{Deutch1968} and Tm\cite{Bernas1973} IPAC experiments have been reported, from which
only the sign information can be reasonably trusted (see e.g. the agreement with other
experiments in Fig.~\ref{fig1}-a and Ref.~\onlinecite{Rao1985}). Due to the the latter sign
information, the HFF of the light lanthanides is guessed to be positive, while for the
heavy lanthanides it is negative. The line in Fig.~\ref{fig1}-a summarizes the most likely
interpretation of this data set.

\begin{figure}
 \begin{center}
  \includegraphics[width=8cm,angle=0]{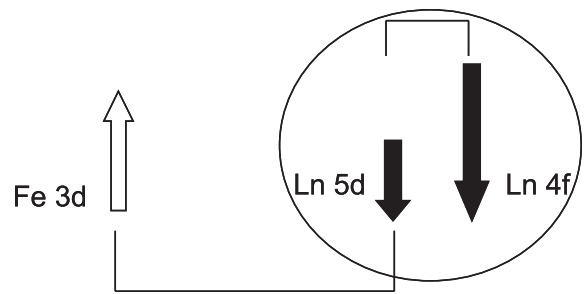}
  \caption{Schematic summary of the model of Campbell and Brooks. The localized 4f spin moment does not interact
           directly with the moments of neighboring atoms. But it does interact with the 5d moment on the same atom
           by ferromagnetic intra-atomic exchange. The more delocalized 5d spin moment interacts by antiferromagnetic
           interatomic exchange with the Fe 3d moment. As a result, the net interaction between Fe-3d and lanthanide-4f is
           antiferromagnetic. A consequence of the model of Campbell and Brooks is the positive (negative) HFF in the first
           (second) half of the lanthanide-series in Fe (Fig.~\ref{fig1}-a). Due to Hund's third rule, the lanthanide
           4f orbital moment is antiparallel (parallel) to the 4f spin moment in the first (second) half of the series.
           Because the orbital contribution to the HFF (which is parallel to the orbital moment) is dominant for
           lanthanides, the total HFF is parallel (antiparallel) to the Fe moment -- and hence called positive
           (negative) -- in the first (second) half of the series. \label{fig2}}

 \end{center}
\end{figure}

\begin{figure}
 \begin{center}
  \includegraphics[width=8cm,angle=0]{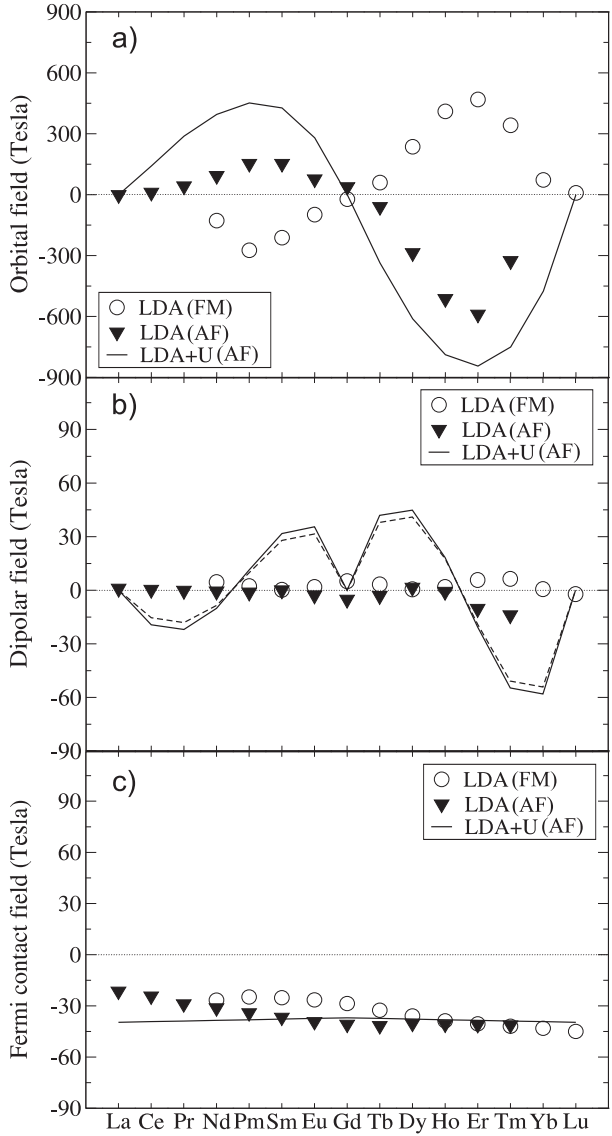}
  \caption{White circles indicate the ferromagnetic solution (FM), black triangles the antiferromagnetic solution (AF).
           Symbols are LDA results, lines are LDA+U results (obtained from Hund's rules occupations, see text).
           Mind the scale, which is 10 times larger in a) compared to
           b) and c). a)~Orbital contribution to the HFF at the lanthanide site due to 4f electrons. b)~Dipolar
           contribution due to 4f only (fully line), and due to 4f
           and 5p (dashed line, see text).
           c)~Fermi contribution. The total HFF is the sum of these 3, and is almost undistinguishable from
           a).\label{fig4}}
 \end{center}
\end{figure}

Fig.\ref{fig1}-a can be understood in terms of Hund's rules and
the model of Campbell and Brooks. Based on heuristic arguments
(Ref.~\onlinecite{Campbell1972}) and first principles calculations
(Ref.~\onlinecite{Brooks1989}), Campbell and Brooks showed that
the \emph{interatomic} exchange interaction between a transition
metal 3d spin moment and a lanthanide 5d spin moment is
antiferromagnetic (Fig.~\ref{fig2}).
The lanthanide 4f moment is localized at the lanthanide site and cannot directly interact
with its transition metal neighbors, but it has a ferromagnetic \emph{intra-atomic}
exchange interaction with the lanthanide 5d moment. The result is a net antiferromagnetic
coupling between the lanthanide 4f moment and the transition metal 3d moment
(Fig.~\ref{fig2}). According to Hund's third rule, the lanthanide orbital moment is
antiparallel to the lanthanide spin moment for the 7 lightest lanthanides, and parallel to
it for the 7 heaviest lanthanides. The dominant contribution to the HFF is the orbital HFF
(see Sec.~\ref{sec-LDA} and Fig.~\ref{fig4}),
which is parallel to the orbital moment. Therefore, one expects the total HFF to be
parallel to the Fe magnetization (and hence positive) for the light lanthanides, and
antiparallel (negative) for the heavy ones, as is seen indeed in Fig.~\ref{fig1}-a.

\subsection{LDA calculations}

\label{sec-LDA} 
\begin{figure}
 \begin{center}
  \includegraphics[width=8cm,angle=0]{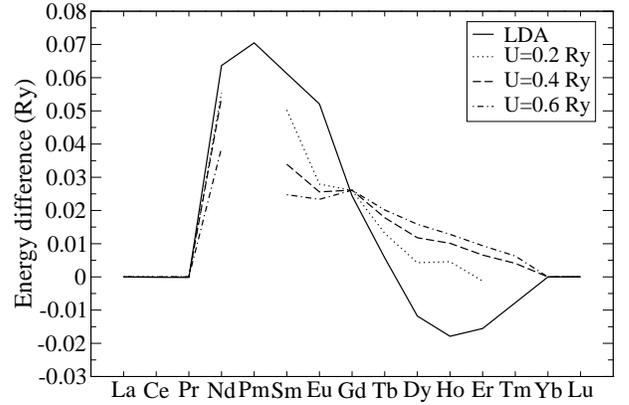}
  \caption{Energy difference between the situations with the lanthanide 4f spin moment ferromagnetically aligned
           with the Fe 3d moment and with the 4f moment antiferromagnetically aligned (E$_{\mathrm{FM}}$ - E$_{\mathrm{AF}}$).
           If this energy difference is positive, the antiferromagnetic situation has the lowest energy. Energy difference
           for LDA are compared with energy differences for LDA+U with progressively larger U. The LDA result is
           equivalent to U=0.0 Ry. There are some gaps in the picture, because not for every situation a converged
           solution could be found. For Pm the calculated energy difference is much larger (0.42~Ry for U=0.6~Ry) and is out
           of scale. \label{fig3}}
 \end{center}
\end{figure}

As a first step, we calculate the magnetic HFF with the common LDA. This
will provide us with a data set to which we can later compare the possible improvement by
LDA+U, and it allows to introduce some peculiarities that will play a role in all later
calculations as well. As is usual with this type of methods, our calculations involve an
iterative procedure (`self-consistent field' procedure) that yields in the end a possible
state of the calculated system, which is not necessarily the desired ground state: in the
space of possible solutions, this self-consistent field procedure finds a \emph{local}
minimum, but not necessarily the \emph{global} minimum. The local minimum that is obtained,
depends to some degree on the starting configuration that was initially chosen. This
behavior is prominently present for lanthanides in Fe. If the spin moment of the lanthanide
initially is put parallel to the Fe spin moment, then this parallelism is maintained
throughout the iterative procedure (except for La, Ce and Pr, where the moment always
spontaneously turns to an antiparallel orientation). We call this from now on the
ferromagnetic solution. With an initially antiparallel configuration, an antiparallel (or
antiferromagnetic) solution is obtained. If the lanthanide was given initially no spin
moment, then a solution with a spin moment that is much reduced compared to the two
preceding solutions was found. In order to decide which of those is the ground state, one
has to look at the total energy of each solution. The total energy of the case with reduced
moment was much higher than the others, and we will not consider it further. The energy
differences between the other two solutions are given in Fig.~\ref{fig3}.
For all lanthanides up to Tb, the antiferromagnetic solution has the lower energy. Starting
with Dy, the ferromagnetic solution becomes the ground state.

In Fig.~\ref{fig4}, the different contributions to the magnetic HFF are given for both
types of solutions. A HFF is a magnetic field at the position of the nucleus, and it is
built mainly from 3 contributions: the spin dipolar field, the Fermi contact field and the
orbital field. The spin dipolar field is generated by the spin moments of the electrons
surrounding the nucleus. For cubic point symmetry, this contribution vanishes. The Fermi
contact field\cite{Fermi1930} is of dipolar nature as well, but is due to the penetration
of s-electrons into the nucleus. It does not vanish for cubic symmetry, and it is the
dominant (and almost only) contribution for impurities up to Z=55 in Fe. The orbital field
stems from the electric charge of the electron that orbits the nucleus, and it vanishes for
cubic symmetry. As Fig.~\ref{fig4} shows, the dipolar field does not exceed a few Tesla,
while the Fermi contact field lies between -20~T and -40~T. The orbital field is the
dominant contribution, and can reach almost $\pm 600$~T. At first sight, one would expect a
zero orbital and dipolar field for a substitutional impurity in Fe, as the point symmetry
is cubic. The reason why this is not the case for lanthanides, is purely due to the
presence of spin-orbit coupling, which breaks the crystalline cubic point group symmetry.
The oscillatory behavior of the orbital moment reflects Hund's third rule: the orbital
moment is antiparallel (parallel) to the spin moment in the first (second) half of the
lanthanide series. Because the orbital field is parallel to the orbital moment, it will for
the antiferromagnetic solution be positive in the first half of the series and negative in
the second half (and vice versa for the ferromagnetic solution).

According to the LDA total energies, we have to accept the
antiferromagnetic solution as the ground state up to Tb, and the
ferromagnetic solution starting from Dy. This leads to positive
HFF's for almost all lanthanides (Fig.~\ref{fig1}-b), which is
contradiction with the current interpretation of the experimental
data set and with the model of Campbell and Brooks. We will
demonstrate in Sec.~\ref{sec-ldau} that this is a new, clear
example of a failure of LDA, to be added to the list of notorious
shortcomings as the general overbinding behavior, the prediction
of the wrong crystal structure for Fe, and the prediction of
metallicity for some strongly correlated insulators as NiO and
La$_2$CuO$_4$. Even if we would select the antiferromagnetic
solution throughout (as the Campbell-Brooks model suggests), then
still the quantitative agreement with the experimental HFF's is
rather poor (Fig.~\ref{fig1}-b).

\subsection{LDA+U calculations}

\label{sec-ldau} Using and interpreting LDA+U calculations brings some complications that
are absent for LDA.  First, LDA+U schemes are not fully \emph{ab initio}: they involve an
on-site Coulomb repulsion parameter $U$ and an on-site exchange interaction constant $J$
that have to be chosen \emph{a priori} for every atom with strong correlations. In our case
we have to choose one $U$ and one $J$ for the f-states of the lanthanide impurity. In line
with the strategy adopted for the relaxation (Sec.~\ref{sec-compdetails}), we strive for
reasonable overall agreement and do not focus on agreement for individual cases too much.
Therefore we take the same $U$ and $J$ for all lanthanides. Looking at other
calculations\cite{Anisimov1991,Harmon1995} and experiments\cite{Herbst1978,Min1986},
$U=0.6~$Ry is a reasonable choice. The value of $J$ is usually an order of magnitude
smaller, and it does not affect the results as much as $U$ does. Therefore we take $J$ as
10\% of $U$. Another -- second -- complication with LDA+U is that due to the introduction
of tuneable variables $U$ and $J$, the total energy looses its exact mathematical meaning
as a variational quantity. If LDA+U is used, a lower energy for one solution does therefor
not necessarily mean that this solution is to be preferred over another one with a higher
energy. This does not mean that these total energies are meaningless: for instance, LDA+U
has been used successfully to determine the equilibrium volume of materials by energy
minimization\cite{Bouchet2000}. The current general opinion is that if the electronic
structure does not \emph{qualitatively} change between two slightly different situations,
then the total energies obtained by LDA+U for these situations can probably be compared.
The criterion to decide whether there is a qualitative difference between two electronic
structures or not, is provided by the 4f density matrix: in qualitatively similar cases,
the occupation of the 14 m-orbitals (= the diagonal elements of the two spin-polarized
$7\times 7$ density matrices) should be more or less identical. Finally -- and third --
there are much more local minima in the space of solutions when LDA+U is used, and a
calculation gets easily trapped in one of them. Because the density matrices for two
different solutions are necessarily different from each other, the total energy cannot be
used to determine which solution is the ground state. This problem is illustrated in
Tab.~\ref{tab1},
\begin{table*}
 \begin{center}
 \caption{Diagonal elements of the 7$\times$7 4f-up density matrix for Tm in Fe (antiferromagnetic case),
          together with the orbital (4f), dipolar (4f+5p) and Fermi contributions to the total HFF (Tesla). These diagonal
          elements give the occupation of each m-orbital (between 0 and 1). The LDA
          result is compared with several LDA+U calculations, all with U=0.6~Ry. The LDA+U calculations differ
          only in the initial distribution of the f-electrons over the different orbitals. In the second column,
          the total energy (in mRy/atom) of the LDA+U calculations is given, relative to case 2, which has the
          lowest energy (see text for discussion). \label{tab1}}
  \begin{ruledtabular}
   \begin{tabular}{lcccccccccccc}
      & $\Delta E$ & m=-3 & m=-2 & m=-1 & m=0 & m=1 & m=2 & m=3 & B$_{\mathrm{orb}}$ &  B$_{\mathrm{dip}}$ & B$_{\mathrm{Fermi}}$ & B$_{\mathrm{tot}}$     \\ \hline
    LDA &  & 0.98 & 0.95 & 0.96 & 0.71 & 0.86 & 0.52 & 0.54 & -326 &
    -14 & -41 & -381 \\
    case 1 & 1.7 & 1.00 & 0.99 & 0.99 & 0.99 & 0.99 & 0.01 & 0.01 & -718
    & -63 & -41 & -822 \\
    case 2 & 0.0 & 0.99 & 0.99 & 0.99 & 0.99 & 0.01 & 0.99 & 0.01 & -571
    & -27 & -41 & -639 \\
    case 3 & 1.6 & 0.99 & 0.99 & 0.99 & 0.02 & 0.99 & 0.99 & 0.01 & -425
    & -16 & -42 & -483 \\
    case 4 & 3.2 & 0.99 & 0.99 & 0.02 & 0.99 & 0.99 & 0.99 & 0.01 & -275
    & -28 & -42 & -345 \\
    case 5 & 8.9 & 1.00 & 0.04 & 0.99 & 0.99 & 0.99 & 0.99 & 0.01 & -126
    & -63 & -42 & -231
    \end{tabular}
  \end{ruledtabular}
 \end{center}
\end{table*}
where the diagonal elements of the f-up density matrix is given for various
antiferromagnetic solutions for Tm in Fe. The 5 electrons can be distributed in different
ways over the 7 orbitals, and always a converged solution can be obtained. The HFF field
can be very different for all cases. In the second column, the total energy of these 5
solutions is given, relative to the case with the lowest energy (case 2). Case 1 -- which
we will later identify as the most probable ground state -- is only third in rank if the
total energy is considered: an illustration of the lack of meaning of the LDA+U total
energy for cases with different density matrices. This makes it impossible to proceed in a
straightforward way as was done for LDA.

Faced with this problem, we first take one step back and examine the energy difference
between the ferromagnetic and the antiferromagnetic case \emph{for the same type of
solution} (i.e.\ for a ferro- and and antiferromagnetic solution that have a similar
f-electron density matrix). These energy differences can be expected to be meaningful (see
above), a hope which is supported by the fact that we find them not to depend on the type
of solution for which we make the comparison. The result is shown in Fig.~\ref{fig3}, for 3
different values of $U$: 0.2, 0.4 and 0.6~Ry (some gaps are present, because not for every
lanthanide a solution could be found for every $U$). Clearly, the use of a non-zero $U$
makes the antiferromagnetic case more stable, for all lanthanides. This brings the sign of
the HFF in agreement with experiment and with the model of Campbell and Brooks. We conclude
that LDA+U describes the effective d-f exchange interaction much better than LDA does, and
that LDA is qualitatively wrong in this respect.

Next, we try to find a way to obtain ground state values for the HFF in the
antiferromagnetic case. To this end, we turn the annoying freedom of having several ways to
occupy the m-orbitals into an advantage, by determining the \emph{individual contribution}
of each m-orbital to e.g. the orbital field. This can be done by first calculating the
orbital HFF for several different antiferromagnetic solutions, as is given as an example
for Tm in Tab.~\ref{tab1}. Then a system of linear equations is set up, with as 7 variables
$x_m$ the orbital HFF of each of the 7 m-orbitals. The occupation of each of these orbitals
(or the diagonal elements of the density matrix from Tab.~\ref{tab1}) are the coefficients.
The occupation found in the calculation should give the calculated orbital field, as is
illustrated here for `case 1' in Tab.~\ref{tab1}:
\begin{eqnarray}
1.00 x_{-3} + 0.99 x_{-2} + 0.99 x_{-1} + 0.99 x_0 + & & \nonumber
\\ 0.99 x_1 + 0.01 x_2 + 0.01 x_3 & = & -718 \nonumber
\end{eqnarray}
This system of equations can be supplemented by other equations expressing some general
truths (the orbital field with all 7 m-orbitals filled is zero, the contribution by +m is
opposite to the one by -m), such that the system becomes overdetermined. Each subset of 7
independent equations should give the same $x_m$, which indeed they do.

\begin{table}
 \begin{center}
 \caption{Overview whether $\mu_{\mathit{orb}}$, B$_{\mathit{orb}}$, B$_{\mathit{dip}}$ and V$_{\mathit{zz}}$
 depend on the shape of the 4f orbital (given by the absolute value of $m$), the direction of motion of
 the electron in that orbital (given by the sign of $m$), and whether they depend on the charge or on
 the spin of the 4f electron in a given $m$-orbital \label{tab3}}
  \begin{ruledtabular}
   \begin{tabular}{l|c|c|c|c}
       & $\mu_{\mathit{orb}}$ & B$_{\mathit{orb}}$ & B$_{\mathit{dip}}$ & V$_{\mathit{zz}}$       \\ \hline
    shape of orbital & yes & yes & yes & yes    \\
  direction of motion &   yes & yes &  no &   no  \\
  charge or spin &  charge & charge &  spin &  charge   \\
    \end{tabular}
  \end{ruledtabular}
 \end{center}
\end{table}

This was done for 10 elements from the lanthanides series (Ce, Nd, Pm, Sm, Tb, Dy, Ho, Er,
Tm and Yb), not only for the orbital HFF but also for the orbital moment and the dipolar
HFF. The dominant contribution to the orbital HFF is due to the 4f electrons, as one could
expect (Fig.~\ref{figB}-a,b). Orbitals with opposite $m$ quantum number yield opposite
orbital hyperfine fields. The latter can be understood as follows (see also
Tab.~\ref{tab3}): in orbitals with opposite $m$, the electrons move in opposite directions,
because opposite $m$ (z-component of the orbital angular momentum) mean that the angular
momenta of those orbitals have different orientations. Hence, the orbital fields will be
opposite as well. As a function of Z, the contribution due to each m-orbital increases. A
linear fit is possible (full line). In order to verify whether this is accidental or not,
we did the same calculations for free lanthanide atoms and free lanthanide 3$^+$ ions. For
the free ions, almost the same perfect linear correlation was found as for the solid
(dotted line in Fig.~\ref{figB}). For free neutral atoms, the fields were slightly larger
(at most 10\% for the orbital field, and 5\% for the dipolar field).
\begin{figure}
 \begin{center}
  \includegraphics[width=8cm,angle=0]{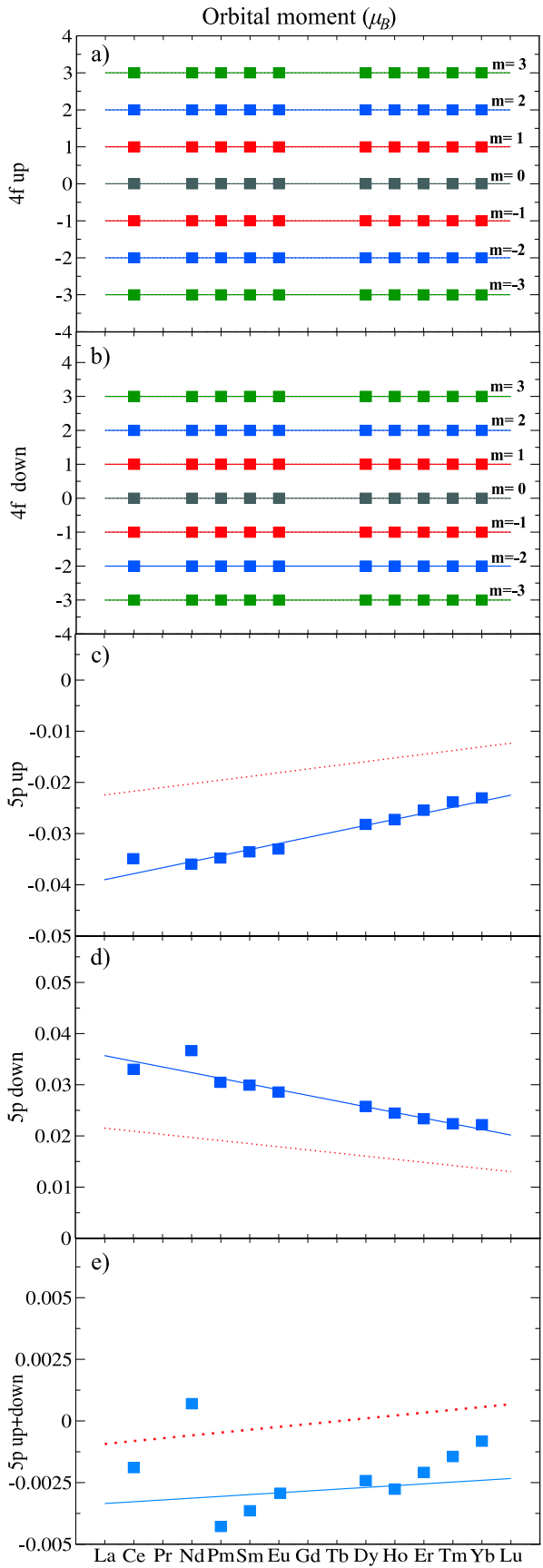}
  \caption{Contribution of each m-orbital to the (a) 4f-up, (b) 4f-down orbital moment and the
  contribution of the (c) 5p-up, (d) 5p-down, (e) 5p up+down electrons to the orbital magnetic moment of a lanthanide
           in Fe. Data points: results from calculations for lanthanides in Fe. Full lines:
           linear fit through these data points. Dotted lines: linear fit through a complete set of calculations
           for free lanthanide ions.  \label{figA}}
 \end{center}
\end{figure}

\begin{figure}
 \begin{center}
  \includegraphics[width=8cm,angle=0]{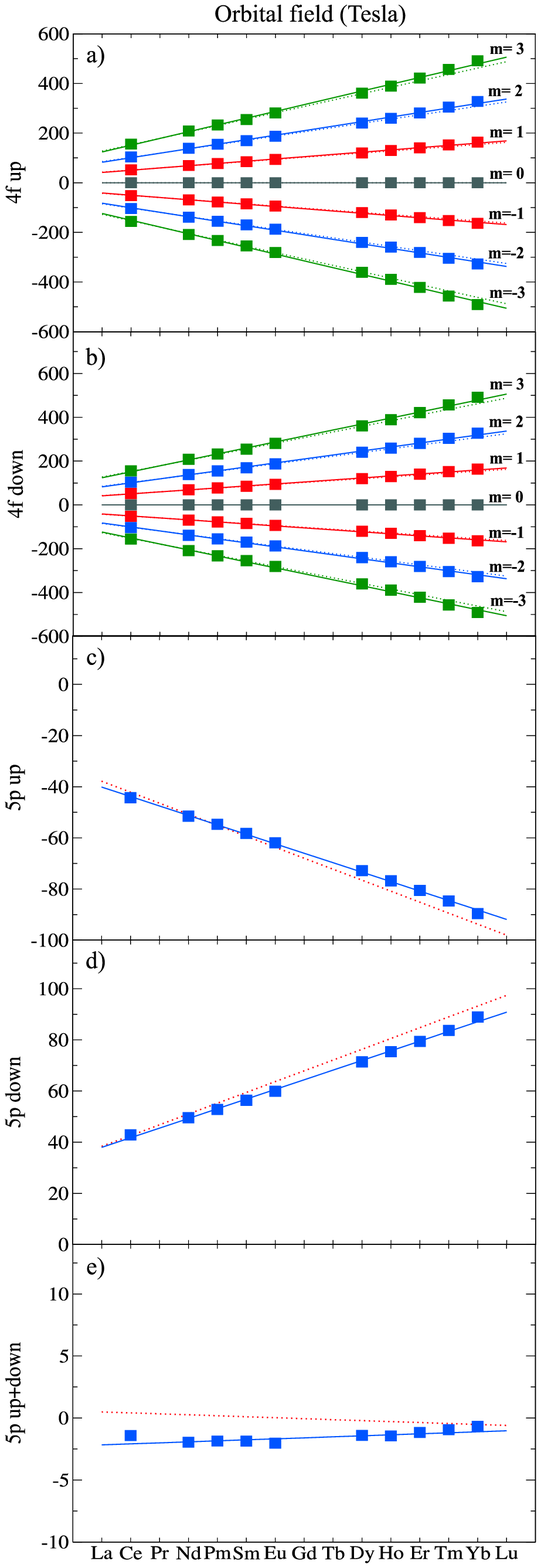}
  \caption{Contribution of each m-orbital to the (a) 4f-up, (b) 4f-down orbital HFF and the
  contribution of the (c) 5p-up, (d) 5p-down, (e) 5p up+down electrons to the orbital HFF of a lanthanide
           in Fe. Data points: results from calculations for lanthanides in Fe. Full lines:
           linear fit through these data points. Dotted lines: linear fit through a complete set of calculations
           for free lanthanide ions.  \label{figB}}
 \end{center}
\end{figure}

\begin{figure}
 \begin{center}
  \includegraphics[width=8cm,angle=0]{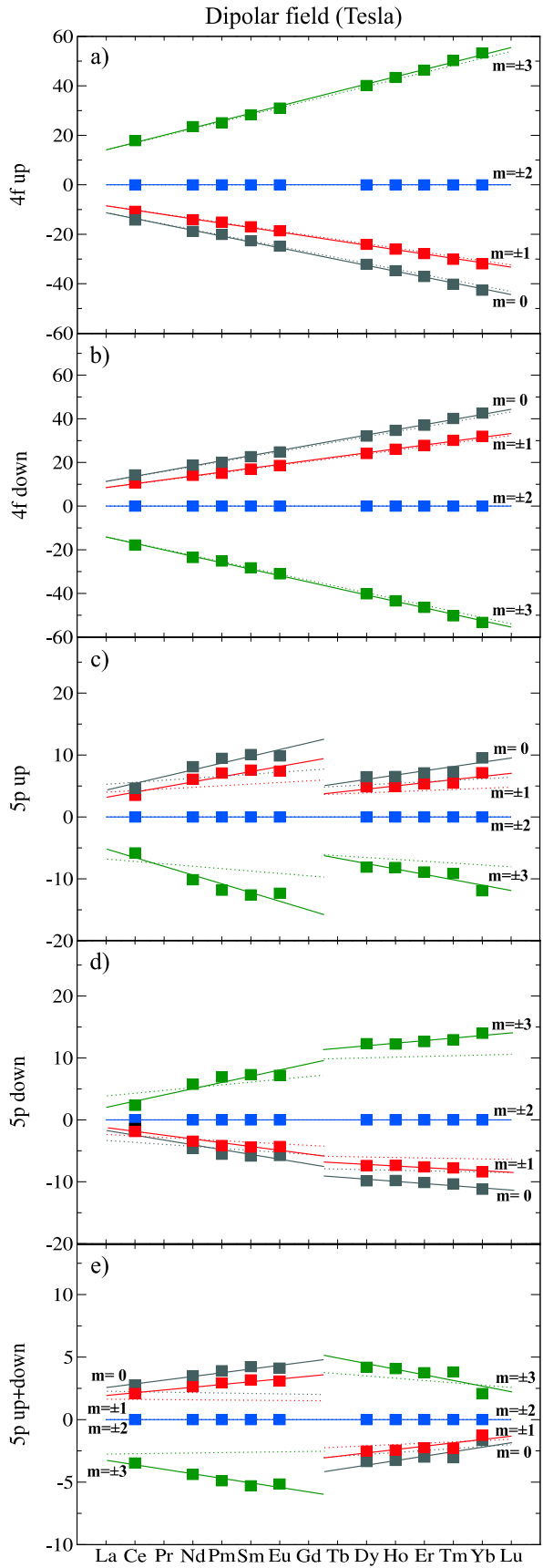}
  \caption{Contribution of each m-orbital to the (a) 4f-up, (b) 4f-down dipolar HFF and the induced (c) 5p-up, (d) 5p-down,
  (e) 5p up+down contributions to the  dipolar HFF of a lanthanide
           in Fe. Data points: results from calculations for lanthanides in Fe. Full lines:
           linear fit through these data points. Dotted lines: linear fit through a complete set of calculations
           for free lanthanide ions.  \label{figC}}
 \end{center}
\end{figure}

\begin{figure}
 \begin{center}
  \includegraphics[width=8cm,angle=0]{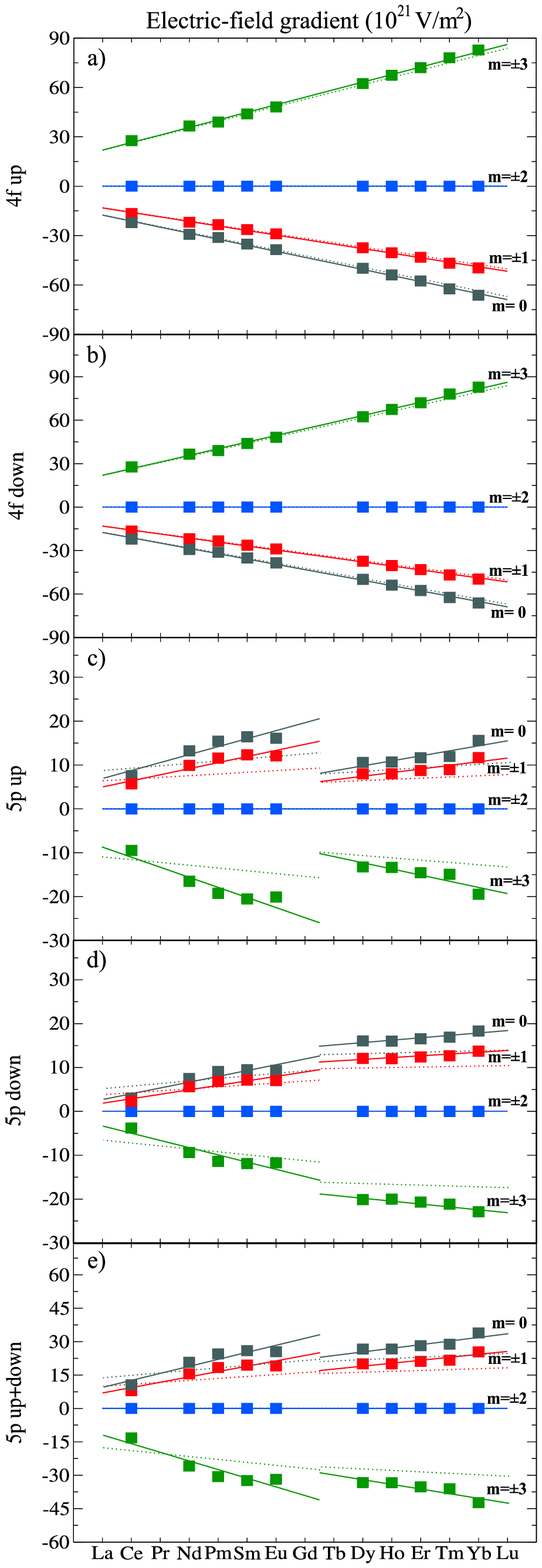}
  \caption{Contribution of each m-orbital to the (a) 4f-up, (b) 4f-down to the total V$_{zz}$ and the induced (c) 5p-up, (d) 5p-down,
  (e) 5p up+down to the total V$_{zz}$ of a lanthanide
           in Fe. Data points: results from calculations for lanthanides in Fe. Full lines:
           linear fit through these data points. Dotted lines: linear fit through a complete set of calculations
           for free lanthanide ions.  \label{figD}}
 \end{center}
\end{figure}
Additional to this large 4f contribution, there is also a 5p contribution to the orbital
HFF (Fig.~\ref{figB}-c-e). The up and down contributions are quite large (40-90 T), but
they cancel each other, yielding a negligible ($<3$~T) total contribution for the
5p-orbital HFF. This 5p-contribution to the orbital HFF does not depend on the
f-configuration. Taking Dy as an example -- 2 electrons in the unfilled spin channel -- the
total 5p orbital field will be -2~T (72~T for 5p-up, -74~T for 5p-down), irrespective
whether these 2 electrons are e.g.\ in the m=+3 and m=+2 or in the m=-1 and m=0 orbitals.
This will be different for dipolar hyperfine fields and for the electric-field gradient.
Looking separately to the 5p up and down contributions, one can see that they don't vanish
for La (4f empty) and Lu (4f full), and that they increase with Z. Moreover, we observed
that this contribution disappears if the SO-coupling is switched off. Therefore we can
conclude that these 5p-contributions are due to an intrinsic p-effect, induced by the
SO-coupling on the p-electrons which breaks the cubic symmetry. Finally there is also a
contribution from the valence 6p-electrons, also induced by the SO-coupling, but this
contribution is really small and can be neglected. The numerical stability of this analysis
can be checked by repeating it for the orbital moment (Fig.~\ref{figA}). The contribution
to the orbital moment of each m-orbital should be m by definition, which is found indeed
(Fig.~\ref{figA}-a,b). The 5p-contributions to the orbital moment have the same origin as
in the case of the orbital HFF, namely the SO-coupling.

For the dipolar HFF as well, the 4f-contribution remains the dominant one
(Fig.~\ref{figC}-a,b), but it is one order of magnitude smaller than the orbital HFF. The
systematics are different from the orbital moment and orbital HFF as well
(Fig.~\ref{figC}-a,b and Tab.~\ref{tab3}). First of all, the dipolar HFF does not depend on
the direction of motion of an electron, such that $\pm m$-orbitals yield the same dipolar
HFF. Secondly, $B_{\mathit{dip}}$ depends explicitly on the electron spin, such that an
electron with opposite spin in the same $m$-orbital yields an oppositie field. Furthermore,
we can observe from Fig.~\ref{figC}-c,d,e that the 5p-contributions to $B_{\mathit{dip}}$
depend on the 4f-occupation. If we take again Dy as an example and put the 2 electrons in
the +3 and +2 orbitals (spin up) we get an induced 5p-contribution of 4~T (-8~T for 5p-up
and 12~T for 5p-down), while this is -3~T (4~T for 5p-up and -7~T for 5p-down) if the 2
electrons are in the +1 and -2 orbitals (spin up). Such a 4f-dependence was not present for
$\mu_{\mathit{orb}}$ and $B_{\mathit{orb}}$. In Sec.~\ref{sec-efgLDAU} we will see that
also for the EFG there is such an explicit 4f-dependence, and we will be able to explain
this by the radial dependencies, which are $1/r$ for $\mu_{\mathit{orb}}$ and
$B_{\mathit{orb}}$, and $1/r^3$ for $B_{\mathit{dip}}$ and $V_{\mathit{zz}}$. Another
observation from Fig.~\ref{figC} is that the induced 5p-contributions behave differently in
the first and the second half of the lanthanide series. This suggest a spin-dependent
interaction: In the first half of the series the unfilled f-band is the down band. The
5p-up contribution is large, the 5p-down is smaller. In the second half the unfilled f-band
is the up band. Now the 5p-up contribution is small and 5p-down smaller. Apparently the
4f-electrons induce a larger contribution in the 5p with opposite spin. The 6p-contribution
remains negligible also for the dipolar HFF.

\begin{figure}
 \begin{center}
  \includegraphics[width=8cm,angle=0]{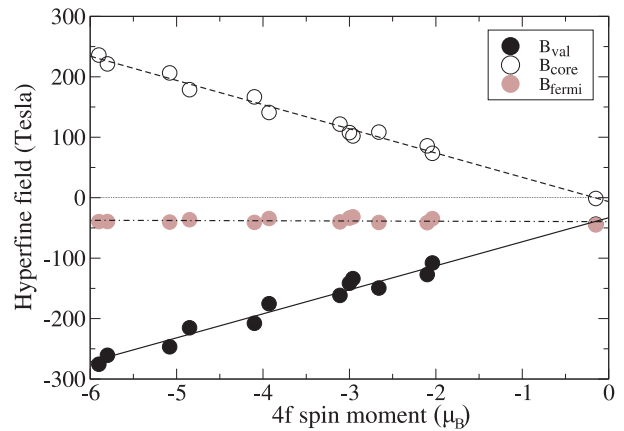}
  \caption{Fermi contact contribution to the HFF as a function of 4f spin moment for a large set of different
           solutions for different lanthanides in Fe. White circles: core contribution. Black circles:
           valence contribution. Gray circles: total Fermi contact contribution. The lines through the core and
           valence contributions are linear fits, the line through the total Fermi field is the sum of those two
           fits.  \label{fig6}}
 \end{center}
\end{figure}

For the Fermi contribution, we cannot follow the procedure of Fig.~\ref{figB}, as this
contribution is less directly connected to the occupation of the f-states. In
Fig.~\ref{fig6},
the Fermi contribution (for valence (mainly 5s and 6s) and core electrons (mainly 1s to 4s)
separately) is plotted as a function of the 4f spin moment, and this for different
lanthanides and different types of solutions. Both the core and valence Fermi contribution
depend linearly on the 4f moment, and they sum to an almost constant value that is
independent of the 4f moment. As this constant contribution is there even for a zero 4f
spin moment, it must be due to other than the 4f electrons. Actually, it results from the
6s-polarization due to the small and relatively constant 5d-moment. The 4f spin moment is
spatially outside the core s-orbitals, but inside the valence s-orbitals. It will therefore
induce exactly the opposite s-polarization on both of them\cite{Watson1961}.

Encouraged by the above observation that the m-orbitals of lanthanides in Fe yield
identical orbital and dipolar fields as for free lanthanide ions, we now suppose that the
ground state occupation of the 4f shell for lanthanides in Fe is the same Hund's rules
ground state as for free lanthanide 3$^+$ ions.

For Tm$^{3+}$ as an example, Hund's rules prescribe the orbitals with m=+3 and m=+2 for the
4f-up spin to be unoccupied, which yields from Fig.~\ref{figB} and Fig.~\ref{figC} the
orbital and dipolar contributions listed in Tab.~\ref{tab0}. The Fermi contact field is
almost independent on the orbital occupation, and is -39~T for Tm and all other lanthanides
in Fe. The total HFF for the ground state of Tm in Fe -- corresponding to a Tm$^{3+}$
configuration -- is hence -745-53-39=-837~T. This value is very close to what was found in
a direct way in `case~1' of Tab.~\ref{tab1}, a solution that has the same type of
occupation as the Hund's rules ground state.

\begin{table}
 \begin{center}
 \caption{ Contributions to B$_{orb}$, B$_{dip}$ and V$_{zz}$ for Tm in Fe in the Hund's rules
 ground state, using the information from figs.~\ref{figB}-\ref{figD} \label{tab0}}
  \begin{ruledtabular}
   \begin{tabular}{l|r|r|r}
      Tm & B$_{orb}$(T) & B$_{dip}$(T) & V$_{zz}$(10$^{21}$V/m$^2$)       \\ \hline
    4f-up & -745 & -49 & -76.2    \\
  4f-down &   -1 &  -1 &   0.9  \\
    5p-up &  -84 &  10 &  16.1   \\
  5p-down &   83 & -13 &  21.2  \\
   6p-up &    10 &   0 &  -0.7   \\
  6p-down &   -8 &   0 &   0.1  \\ \hline
     Sum  & -745 & -53 & -38.6    \\
    \end{tabular}
  \end{ruledtabular}
 \end{center}
\end{table}

\begin{figure}
 \begin{center}
  \includegraphics[width=8cm,angle=0]{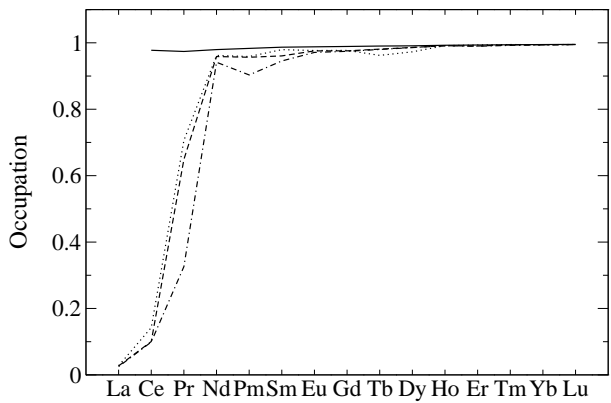}
  \caption{The average occupation of the m-orbitals for the free ions (full line) compared with the occupation of
  the 3 orbitals in the unfilled spin-channel that are most occupied for lanthanide in Fe
  (dotted, dashed and dotted-dashed lines) \label{fig0}}
 \end{center}
\end{figure}

In this way, we can obtain by Fig.~\ref{figB} an LDA+U orbital HFF for all lanthanides --
in the antiferromagnetic type of solution -- which is given by the full line in
Fig.~\ref{fig4}-a. In the same way, Fig.~\ref{figC} leads to the LDA+U dipolar field, given
by the full line in Fig.~\ref{fig4}-b and Fig.~\ref{fig6} to the LDA+U Fermi field in
Fig.~\ref{fig4}-c. (It should be noted here that the core part from this calculated Fermi
field most likely suffers from the typical `LDA core error', for which recently a promising
cure has been proposed\cite{Novak2003}.) The sum of all these contributions leads to an
LDA+U value for the total HFF of all lanthanide 3$^+$ ions in Fe, which is given by the
full line in Fig.~\ref{fig1}-b. The agreement with experiment is rather good for many
lanthanides, especially in the middle of the series. Towards the beginning and the end,
there is moderate disagreement. Only for Yb there is a large deviation between an accurate
and reliable experimental value of -125~T (TDPAC) and a calculated value for the 3$^+$ ion
of -570~T. Yb, however, often occurs in a divalent configuration, where the 4f shell is
completely filled. This results in the -39~T of the Fermi contact field only, which is in
much better agreement with experiment. Nevertheless, a trivalent configuration was
suggested before\cite{Devare1978}, from the following experimental considerations: the
-125~T for Yb in Fe was considered to be `large', much larger than the (-)61~T for the
divalent Lu in Fe. This additional -64~T was taken as stemming from an orbital
contribution, which must lead to the conclusion that the Yb is trivalent. From
Fig.~\ref{fig1}-a, however, we see that such a difference of 64~T is almost negligible, and
that the expected orbital contribution for a trivalent state would be 10 times larger. The
HFF for the other lanthanides in a divalent configuration is shown in Fig.~\ref{fig1}-b as
well (dashed line). Although the agreement is better now for Er and Tm, the uncertainty on
the calculated values is too large to pretend that they would be divalent as well (which is
for Er extremely unlikely, anyway). Another lanthanide that is often divalent is Eu. The
experimental HFF lies halfway between the HFF for Eu$^{2+}$ and Eu$^{3+}$, such that at
this stage no definite conclusion on the valency is possible yet. M\"{o}ssbauer isomer
shift data\cite{Cohen1974,Niesen1978}, however, convincingly point to trivalency for Eu in
Fe (see Sec.~\ref{sec-efgLDAU} for a continuation of this discussion and a final conclusion
on the Eu valency). Irrespective of the valence configuration we consider, one can see that
a moderate deviation from experiment for La, Ce and Pr still remains for the LDA+U
calculations. The LDA results are closer to experiment. As was mentioned in
Sec.~\ref{sec-intro}, LDA calculations for lanthanides represent an itinerant (also called
delocalized) 4f configuration, which is mostly not what is found in Nature: the radius of
the 4f orbitals is not very large, overlap with the orbitals of neighboring atoms is
negligible, and as a result the 4f orbitals are localized. The only exception is Ce. As the
4f radius gets smaller for increasing atomic number $Z$, the 4f orbitals of Ce reach most
outwards. The radius is just large enough to allow overlap and hence delocalization in
materials where the nearest neighbor distance is not too large. Therefore, Ce can be either
trivalent or itinerant, depending on the material\cite{Smith1985,Min1986}. One can say that
in the lanthanide series there is a delocalization-localization transition, that happens
already at the very first element, Ce. The good agreement between experiments and the
LDA-results up to Pr in Fig.~\ref{fig1}-b, suggests that not only Ce but also Pr has
delocalized 4f electrons in an Fe host: a `postponed' delocalization-localization
transition. Three additional arguments support this hypothesis. The first one is given in
Fig.~\ref{fig0}, where for all lanthanides in Fe the number of electrons inside the muffin
tin sphere with radius $R_{\mathit{mt}}=2.45$~a.u.\ is plot for the 3 most occupied
$m$-orbitals in the unfilled spin channel, and compared with the average number of
electrons in the same muffin tin sphere for occupied orbitals in the free lanthanide ions
(which is almost unity). All data are for LDA+U calculations. Even though LDA+U favors
localization, one can see that the occupation of the orbitals for Ce and Pr is far from 1:
the 4f electrons are distributed in noninteger quantities over all 7 $m$-orbitals and/or
reach out of the muffin tin sphere. This is a signature of delocalized behavior. (Two
notes: 1. our procedure to extract the contributions for a single orbital works even with
such noninteger occupations as input. The LDA+U values in Figs.~\ref{fig1}-b and~\ref{fig8}
are therefore really values for localized orbitals, although this localization is not
realized in the LDA+U calculations themselves; 2. Fig.~\ref{fig0} should not be considered
as a proof that from Nd on the 4f electrons are localized. These are results from LDA+U,
which favors localization. Even LDA+U is not able to localize the 4f up to Pr. Therefore
fig.~\ref{fig0} offer a lower bound for the delocalization transition (delocalization
\emph{at least} up to Pr). It does not exclude that, in Nature, Nd, Pm,... in Fe could be
delocalized as well.) A second supporting argument comes from the lanthanide-Fe distance:
2.60~\AA, which is considerably smaller than a typical lanthanide-lanthanide distance in
pure lanthanide metals (4.08~\AA). One can expect from this a large overlap between the 4f
orbitals and the Fe-3d, and hence a stronger tendency to delocalization. This argument will
be further quantified in terms of pressure in Sec.~\ref{sec-pressure}. Finally, in
Sec.~\ref{sec-efgLDAU}, we will show that also the EFG of Ce in Fe indicates delocalized 4f
behavior. In conclusion, we can say that there are several strong indications that the
delocalization-localization transitions for lanthanides in Fe is somewhat postponed, at
least up to Pr. We will come back to this in Sec.~\ref{sec-pressure}.

Regardless of the ambiguity to choose between divalent and trivalent configurations, one
conclusion can be made unambiguously from Fig.~\ref{fig1}-b: the magnitudes of the HFF's
with LDA+U are much closer to experiment than with LDA, certainly for the heavier
lanthanides. Together with the antiferromagnetic coupling which is reproduced by LDA+U but
not everywhere by LDA (Fig.~\ref{fig3}), this is hard evidence for the fact that LDA+U
performs considerably better than LDA also in these systems.

\section{Electric-Field Gradients}
\label{sec-efg}
\subsection{Experimental data set}
\begin{figure}
 \begin{center}
  \includegraphics[width=8cm,angle=0]{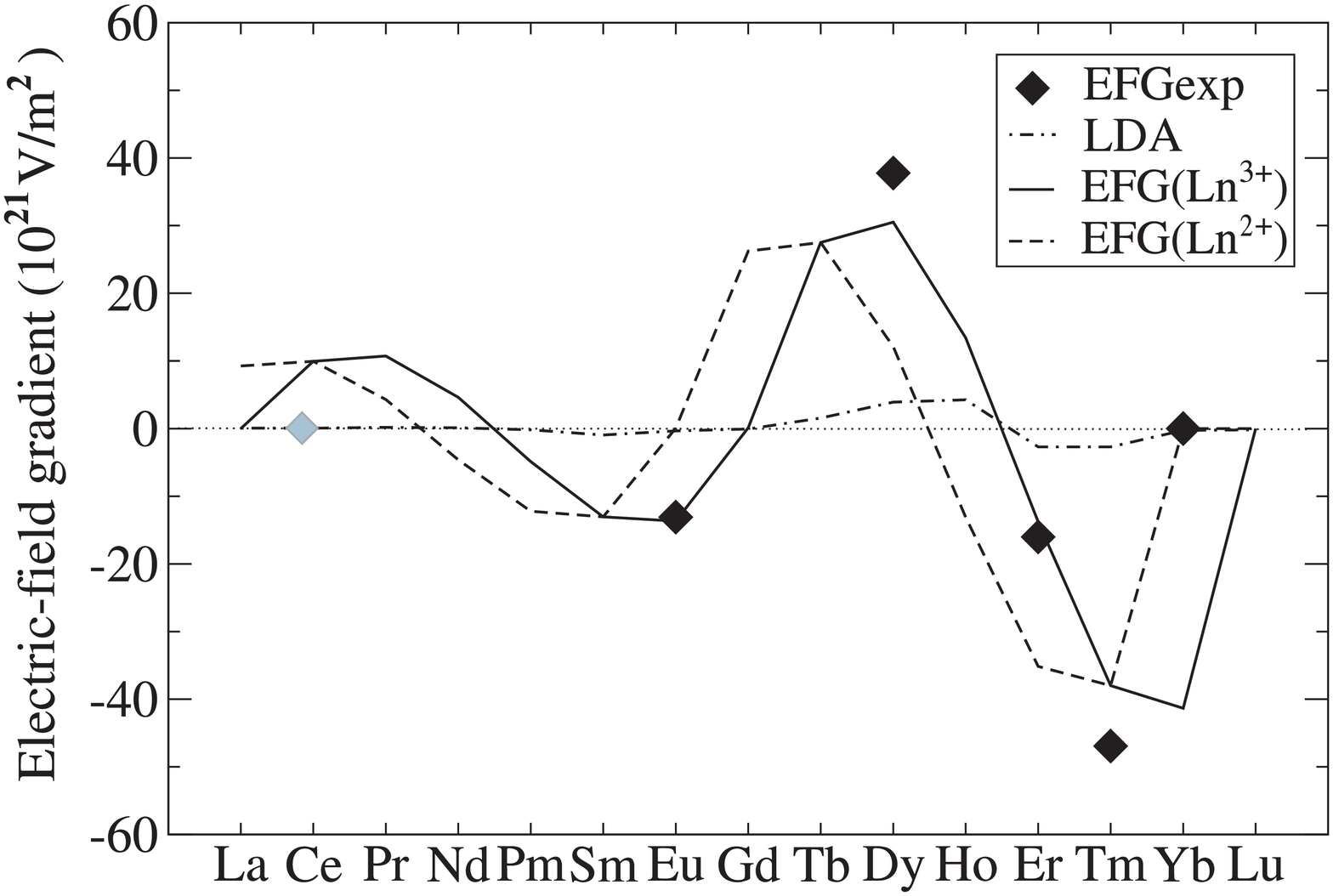}
  \caption{$V_{\mathrm{zz}}$ for lanthanides in Fe. Symbols: experimental
           data\cite{Niesen1976,Niesen1978,Devare1978,VanRijswijk1982} (see text).
           The gray symbol for Ce indicates that this value is a -- rather safe -- guess (see text).
           Dotted-dashed line: LDA results.
           Full line: LDA+U values for trivalent lanthanides (using the information of Fig.~\ref{figD}, see text).
           Dashed line: LDA+U values for divalent lanthanides.
           \label{fig8}}
 \end{center}
\end{figure}

Only few experimental data on the main component $V_{\mathrm{zz}}$
of the electric-field gradient tensor for lanthanides in Fe are
available (Fig.~\ref{fig8}).
For Eu, Dy, Er and Tm, M\"{o}ssbauer measurements\cite{Niesen1976,Niesen1978} were done.
Some attention is needed for Eu, which is reported to have $V_{\mathrm{zz}}=0$ in
Ref.~\onlinecite{Niesen1976}, based on $^{151}$Eu M\"{o}ssbauer spectroscopy from
Ref.~\onlinecite{Cohen1974}. Niesen and Ofer, however, have later shown\cite{Niesen1978} by
$^{153}$Eu M\"{o}ssbauer spectroscopy that $V_{\mathrm{zz}}=-11.1\,10^{21}$~V/m$^2$. For Ce
in Fe the quadrupole coupling constant (which contains the product between
$V_{\mathrm{zz}}$ and the quadrupole moment Q) has been determined by $^{141}$Ce
NMR\cite{VanRijswijk1982} to be almost zero. The quadrupole moment for $^{141}$Ce is not
known, but assuming a typical value of 1 barn leads to a $V_{\mathrm{zz}}$ that is
practically zero (gray symbol in Fig.~\ref{fig8}). The zero $V_{\mathrm{zz}}$ for Yb is not
explicitly mentioned in the literature, but can be inferred from Fig.~1 in
Ref.~\onlinecite{Devare1978}, which shows a purely magnetic interaction.

\subsection{LDA+U calculations}

\label{sec-efgLDAU} As in an LDA-calculation all f-orbitals are roughly equally populated,
there is almost no spatial anisotropy and $V_{\mathrm{zz}}$ will be close to zero
(dotted-dashed line in Fig.~\ref{fig8}). We therefore turn immediately to LDA+U
calculations. We are confronted with the same problems as described in Sec.~\ref{sec-ldau}
-- $V_{\mathrm{zz}}$ sensitively depends on the f-electron density matrix and there is no
criterion to determine which density matrix corresponds to the ground state -- and
therefore we determine again the individual contribution to $V_{\mathrm{zz}}$ of every 4f
m-orbital and the induced 5p contributions to $V_{\mathrm{zz}}$ (Fig.~\ref{figD} and
Tab.~\ref{tab3}). Just as for the dipolar HFF, the EFG due to the 4f orbitals themselves
does not depend on the direction of motion of the electron, and $\pm m$-orbitals yield the
same $V_{\mathit{zz}}$. On the other hand, the EFG depends on the charge and not on the
spin, such that also up and down 4f-electrons yield the same $V_{\mathit{zz}}$. The 5p
contribution depends on the 4f occupation, just as for the dipolar HFF. Assuming a Hund's
rules type of occupation either for divalent or trivalent ions, leads to reasonable
agreement with experiment (Fig.~\ref{fig8}). Fig.~\ref{fig8} provides further evidence for
the fact that Yb in Fe really is divalent: trivalent Yb has a very large $V_{\mathrm{zz}}$,
while the experimental value is zero, in agreement with the divalent prediction. For Eu,
the measured negative $V_{\mathrm{zz}}$ is in good agreement with the calculated value for
a trivalent state. Together with the experimental evidence based on the isomer shift
(Sec.~\ref{sec-ldau}), we can now firmly conclude that Eu in Fe is indeed trivalent, and
has a valence state which is different from the one of Yb. For Er, Fig.~\ref{fig8} suggests
trivalency as well, a conclusion that could not be unambiguously made based on the HFF only
(Sec.~\ref{sec-ldau}). For Tm, the situation remains undecided. Irrespectively of the
valence state we consider for Ce, a large deviation from experiment can be observed.
However, the experiment matches very well with the itinerant LDA result. This is another
indication that Ce in Fe has delocalized f-electron.

\begin{figure}
 \begin{center}
  \includegraphics[width=7cm,angle=0]{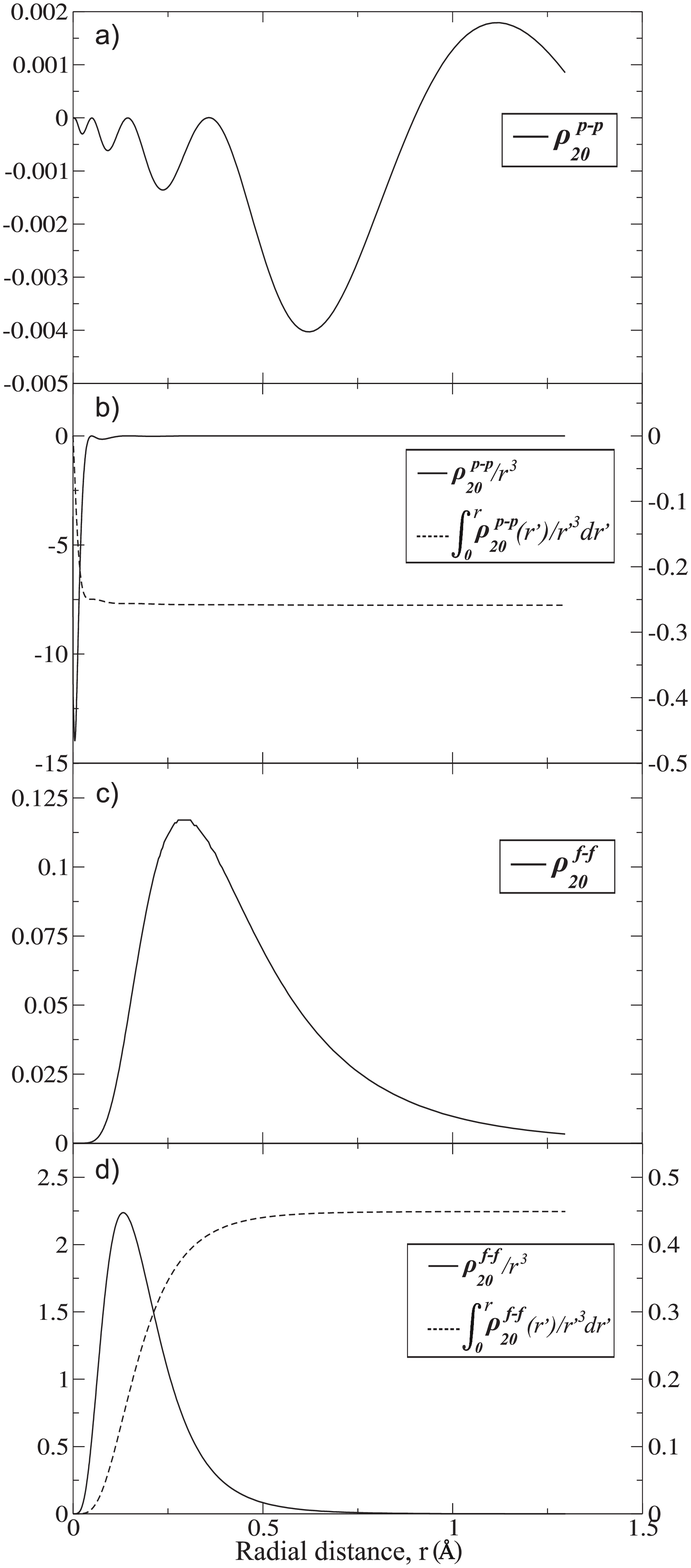}
  \caption{a) The anisotropic p-density $\rho^{\mathrm{p-p}}_{20}(r)$ for Tb in Fe (sum of up and down electrons,
           arbitrary units for y-scale). b) Left axis (arbitrary units): $\rho^{\mathrm{p-p}}_{20}(r)/r^3$ for
           Tb in Fe. Right axis (arbitrary units): integral of $\rho^{\mathrm{p-p}}_{20}(r)/r^3$,
           which is called $V^{\mathrm{p-p}}_{\mathrm{zz}}$ (apart from a constant factor with negative sign).
           c) and d): idem, but for the f-f contribution.
           \label{fig11}}
 \end{center}
\end{figure}

Now we can analyze which are the main electrons that provide the anisotropy that leads to
the EFG. It has been shown before\cite{Blaha1988} in a rigorous way that for metals with
s-, p- and d-electrons the total $V_{\mathrm{zz}}$ can be obtained as a sum of a quantity
$V^{\mathrm{p-p}}_{\mathrm{zz}}$ and $V^{\mathrm{d-d}}_{\mathrm{zz}}$ (neglecting small
contributions from the interstitial region of the crystal). They measure the nonspherical p
and d charge densities $\rho^{\mathrm{p-p}}_{20}(r)$ and $\rho^{\mathrm{d-d}}_{20}(r)$,
respectively, weighted by an integral over $1/r^3$:
\begin{eqnarray}
V^{\mathrm{p-p}}_{\mathrm{zz}} \propto \int_0^R
\frac{\rho^{\mathrm{p-p}}_{20}}{r^3}\,dr \\
V^{\mathrm{d-d}}_{\mathrm{zz}} \propto \int_0^R
\frac{\rho^{\mathrm{d-d}}_{20}}{r^3}\,dr
\end{eqnarray}
R is the radius of the muffin tin sphere of the considered atom. The factor $1/r^3$
strongly emphasizes the contribution from the region close to the nucleus, with small $r$.
s-electrons do not contribute as they have spherical symmetry, and so-called `mixed' s-d or
s-p contributions are negligible and therefor omitted. This can be extended to materials
with f-electrons, such that $V_{\mathrm{zz}}$ for lanthanides can be written as:
\begin{eqnarray}
V_{\mathrm{zz}} \approx V^{\mathrm{p-p}}_{\mathrm{zz}} +
V^{\mathrm{d-d}}_{\mathrm{zz}} + V^{\mathrm{f-f}}_{\mathrm{zz}}
\end{eqnarray}
We now apply this analysis to Tb in Fe, which is a particularly clear example because all
4f-down orbitals are fully occupied and there is only a single 4f-up electron.
Tab.~\ref{tab2} shows the different contributions to $V_{\mathrm{zz}}$ when this single
electron is put in the m=-1 (up) orbital (this is not the ground state, but this is just an
example, anyway).
\begin{table}
 \begin{center}
 \caption{Contributions to $V_{\mathrm{zz}}$ for Tb in Fe, with the down-channel for Tb-4f completely filled
          and with for the 4f-up channel one electron in the m=-1 orbital. The first column gives the rigorous
          notation of each contribution (see Ref.~\onlinecite{Blaha1988}). In the second column an interpretative
          notation is defined, that is used in the text and in Fig.~\ref{figD}. The third column gives the energy
          region in the Density Of States (DOS) near to which these states are found (E$_{\mathrm{F}}$ means ``near
          the Fermi energy'', negative values are below the Fermi energy). Units: 10$^{21}$~V/m$^2$. \label{tab2}}
  \begin{ruledtabular}
   \begin{tabular}{l|l|crrr}
      & & in DOS (eV) & up & down &      \\ \hline
    $V^{\mathrm{d-d}}_{\mathrm{zz}}$ (5d) & $V^{\mathrm{5d}}_{\mathrm{zz}}$ & E$_{\mathrm{F}}$ & 0.1 & 0.4 &  \\
    $V^{\mathrm{p-p}}_{\mathrm{zz}}$ (5p) & $V^{\mathrm{5p}}_{\mathrm{zz}}$ &-23 & 5.6 & 9.6 &  \\
    $V^{\mathrm{p-p}}_{\mathrm{zz}}$ (6p) & $V^{\mathrm{6p}}_{\mathrm{zz}}$ &E$_{\mathrm{F}}$ & -0.2 & 0.4 &  \\
    $V^{\mathrm{f-f}}_{\mathrm{zz}}$ (4f) & $V^{\mathrm{4f}}_{\mathrm{zz}}$ &-5 & -29.0 & 2.5 &  \\  \hline
    $V_{\mathrm{zz}}$  & & & -23.5 & 12.9 & sum = -10.6 \\
    \end{tabular}
  \end{ruledtabular}
 \end{center}
\end{table}
Tab.~\ref{tab2} shows that the main contribution to the total
$V_{\mathrm{zz}}\,=\,-10.6\,10^{21}~\mathrm{V/m}^2$ is due to the single 4f-up electron:
$V^{\mathrm{4f}}_{\mathrm{zz}}\,=\,-29.0$. There is a large contribution of 5.6+9.6=15.2
with the opposite sign due to p-electrons. What is surprising is that this p-contribution
does not stem from the valence 6p electrons, but from the entirely filled and strongly
bound 5p shell, which lies more than 20~eV below the Fermi energy. Intuitively, one would
have assumed such a filled and well-bound shell to be entirely spherically symmetric, which
would mean $V^{\mathrm{5p}}_{\mathrm{zz}}=0$. And indeed, the 5p-anisotropy $\Delta p \,=\,
\frac{1}{2}\left(n_{\mathrm{p_x}} + n_{\mathrm{p_y}}\right) - n_{\mathrm{p_z}}$ is very
small: 0.0030 (up) and 0.0040 (down) (it is shown in Ref.~\onlinecite{Blaha1988} that
$\Delta p$ is proportional to $V^{\mathrm{5p}}_{\mathrm{zz}}$ ; $n_{\mathrm{p_i}}$ is the
number of electrons in the $\mathrm{p_i}$ orbital, and $\Delta p$ measures the unequal
occupation of the 3 p-orbitals). However, a considerable part of this anisotropy stems from
a region very close to the nucleus and hence gets amplified by the $1/r^3$ factor. This is
demonstrated in Fig.~\ref{fig11},
where for the same Tb-configuration the bare anisotropic p-p charge density
$\rho^{\mathrm{p-p}}_{20}(r)$ is shown, before (Fig.~\ref{fig11}-a) and after
(Fig.~\ref{fig11}-b) weighing with a 1/r$^3$ factor, and also after integration
(Fig.~\ref{fig11}-b, right axis). Figs.~\ref{fig11}-c and~\ref{fig11}-d repeat this for the
f-f contribution. The final integrals are clearly determined exclusively by anisotropies in
a region closer than 0.05~\AA\ to the nucleus. In the expression for a spin dipolar field,
the same factor $1/r^3$ is present, explaining why a similar dependence of the 5p dipolar
field on the 4f occupation was observed there (Fig.~\ref{figC}-c,d,e).

A generalization of this analysis is given in Fig.~\ref{figD}, which shows the same
individual 4f m-orbital contributions to $V_{\mathrm{zz}}$ for lanthanides split into 4f
and 5p (i.e. when a given 4f m-orbital is occupied, Fig.~\ref{figD}-a,b shows the direct
contribution from this orbital, while Fig.~\ref{figD}-e shows the corresponding induced
contribution of the 5p-shell (up and down summed)). This shows that the opposite signs for
5p and 4f as seen in the example of Tb is a general effect: occupying the 4f m=0 orbital
gives a negative direct 4f contribution but induces a positive 5p contribution, etc. This
can be understood as follows. A negative $V_{\mathrm{zz}}$ corresponds to charge
accumulation along the z-axis, a positive $V_{\mathrm{zz}}$ to charge accumulation in the
xy-plane. The shape of the 4f orbitals is such that m=0 has its charge mainly along the
z-axis (reflected in a negative $V^{\mathrm{4f}}_{\mathrm{zz}}$), while the xy-plane is
more and more occupied for larger $\left| m \right|$. Apparently the 4f electrons dispel
the 5p electrons: if m=0 is occupied, then the 5p electrons are forced away from the z-axis
into the xy-plane, resulting in a positive $V^{\mathrm{5p}}_{\mathrm{zz}}$ and $\Delta p$.
As Tab.~\ref{tab2} shows, a 4f-up electron distorts the 5p orbitals with either spin: this
is an interaction between the electron charges, not between the spins.

Recently, the EFG of the actinide U has been analyzed in UO$_2$ by R.~Laskowski et al.
(Ref.~\onlinecite{Laskowski2004}), using the same APW+lo method as used in this work. These
authors show in their Fig.~2 the contributions of $V^{\mathrm{p-p}}_{\mathrm{zz}}$,
$V^{\mathrm{d-d}}_{\mathrm{zz}}$ and $V^{\mathrm{f-f}}_{\mathrm{zz}}$ as a function of the
deformation of the oxygen cage that surrounds the U atom. They do not further divide the
p-p contribution in 6p and 7p (here the 6p shell is entirely filled and well-bound).
Without deformation of the oxygen cage, the crystallographic surrounding is cubic and the
slightly non-zero $V_{\mathrm{zz}} \approx -2\,10^{21}~V/m^2$ is due to spin-orbit coupling
only, as is the case for lanthanides in Fe (compare with Nd in Fig.~\ref{fig8}). This small
$V_{\mathrm{zz}}$ is a sum of a f-f contribution of +20 and a p-p contribution of -21 (the
d-d contribution is small: -1). This can be compared to Fig.~\ref{figD} for Nd with the
m=(-3,~-2,~-1) orbitals filled, which leads to a 4f-contribution +14 and a 5p-contribution
of -12, quite similar values. Because of this analogy, we suggest that also for U in UO$_2$
th p-p contribution for an undistorted oxygen cage is due to the completely filled 6p
shell. This interpretation would furthermore imply that as a function of oxygen cage
deformation, the 6p-contribution in Fig.~2 of Ref.~\onlinecite{Laskowski2004} would remain
almost constant (just as the 5f-contribution does), and that the strong decrease of the
total p-p contribution is due to 7p only. This makes sense, as this decrease is
attributed~\cite{Laskowski2004} to the tails of the O-2p wave functions, and hence should
appear near the Fermi energy (= the region of the 7p).

\section{Elaborations}
\subsection{Pressure} \label{sec-pressure}
\begin{figure}
 \begin{center}
  \includegraphics[width=7cm,angle=0]{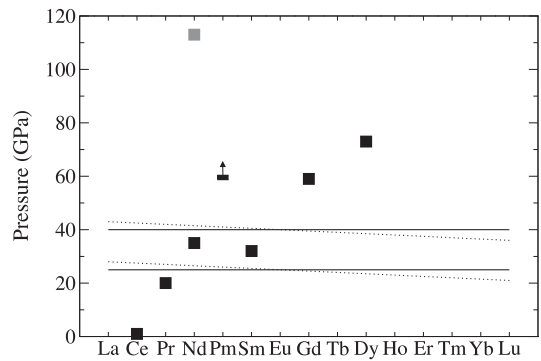}
  \caption{The experimental transition pressure from localized 4f to delocalized 4f
  electrons for pure lanthanides (squares) and two possible choices for the effective pressure felt
  by lanthanides in Fe (full and dotted lines) (for details see text)
           \label{pressure}}
 \end{center}
\end{figure}

It is well known experimentally\cite{McMahan1998} that the pure lanthanides exhibit a large
variety of structural phase transitions as a function of external pressure. At certain
pressures, volume collapses are sometimes observed and attributed to the delocalization of
the f electrons. These delocalization pressures have been determined for 6 elements of the
lanthanide series: Ce~\cite{Zachariasen1977,StaunOlsen1985},
Pr~\cite{Smith1982,Grosshans1983,Zhao1995,Baer2003},
Nd~\cite{Zhao1994,Akella1999,Chesnut2000}, Sm~\cite{Grosshans1984,StaunOlsen1990,Zhao1994},
Gd~\cite{McMahan1998} and Dy~\cite{Patterson2004} (Fig.~\ref{pressure}). For Ce the
delocalization of the f electrons occurs around the pressure of 1~GPa and is accompanied by
a volume collapse of 16\% at the isostructural transition to another fcc phase
($\alpha$-Ce). Pr transforms to a $\alpha$-U structure at 20~GPa with a volume collapse of
9-12\%. In Gd 4f delocalization occurs at 59~GPa when the structure changes to
body-centered monoclinic (bcm) with a volume collapse of 10.6\% and in Dy this happens at
73~GPa with a volume collapse of 6\%. For Nd and Sm no volume collapse has been observed.
In these two cases the delocalization of the 4f electrons was associated with the
appearance of low-symmetry structures (similar with those that appear in Pr, Gd and Dy
cases with volume collapse). This is a somewhat ambiguous procedure. For Nd two transition
pressures have been proposed: 40~GPa (corresponding to the transition to an hP3
structure~\cite{Zhao1994,Akella1999}) and 113~GPa (corresponding to the transition to the
$\alpha$-U phase~\cite{Chesnut2000}). For Sm the delocalization pressure is proposed to be
37~GPa, when the Sm structure changes to hP3. For Pm we have only a lower limit for the
transition pressure, 60~GPa (until this pressure no low-symmetry structure has been
observed\cite{Haire1990}).
In our calculations for lanthanide impurities we replace an Fe atom from an iron lattice by
a lanthanide atom. Obviously the lanthanide atom (which has a much larger volume) will feel
a chemical pressure or effective pressure. How large will this effective pressure be? We
concluded in Sec.~\ref{sec-ldau} that at least Ce and Pr are delocalized. Hence, the
effective pressure -- which we assume to be independent on the lanthanide in a first
approximation -- should be at least 20~GPa (Fig.~\ref{pressure}). The hyperfine fields for
Nd and Sm are only very approximately measured (Fig.~\ref{fig1}-a), such that one cannot
conclude whether they are localized or not. Based on isomer shift and EFG
(Fig.~\ref{fig8}), Eu is definitely localized, as are all heavier lanthanides. Therefore,
two qualitatively different proposals for the effective pressure are possible: about 25~GPa
(everything starting with Nd is localized) or about 40~GPa (everything below Eu is
delocalized, except for Pm and maybe Nd). Assuming an effective pressure that is not
constant (motivated by the decreasing volume of heavier lanthanides) does not change this
picture (dotted lines in Fig.~\ref{pressure} -- these lines qualitatively take the
lanthanide contraction into account). A more accurate experimental determination of HFF and
EFG for Nd, Pm and Sm in Fe would allow to distinguish between both scenarios and would
allow to determine the real position of the delocalization-localization transition in this
system. Experimental data for Nd in Fe are also for another reason interesting: if Nd in Fe
would be found to be localized (itinerant) and the effective pressure of 40~GPa would known
to be correct (from a Sm-measurement for instance) than the delocalization pressure of
113~GPa (35~GPa) for bulk Nd is probably correct. If the effective pressure of 25~GPa would
be correct, no such conclusion can be made.

In conclusion for this section, accurate measurement of HFF and EFG for Nd, Pm or Sm in Fe
would offer a lot of information.

\subsection{Free lanthanide ions}

\begin{figure}
 \begin{center}
 \includegraphics[width=8cm,angle=0]{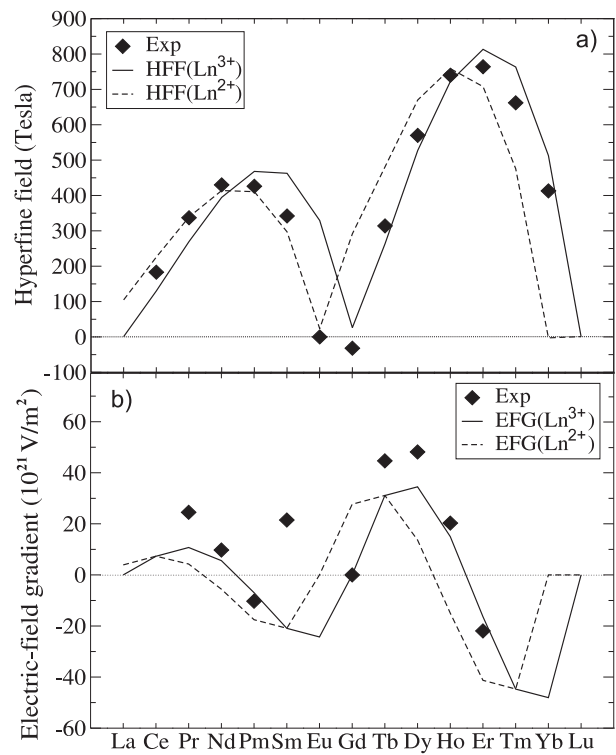}
  \caption{a) Experimental value for the HFF in free lanthanide ions, compared with LDA+U predictions
           based on Figs.~\ref{figB},~\ref{figC} and~\ref{fig6}, both for divalent and trivalent lanthanides.
           b) Experimental value for $V_{\mathrm{zz}}$ in
           free lanthanide ions, compared with LDA+U predictions based on Fig.~\ref{figD}, both for divalent
           and trivalent lanthanides. \label{fig10}}
 \end{center}
\end{figure}

LDA+U calculations for free lanthanide 3$^+$ ions were already mentioned in
Figs.~\ref{figA}-~\ref{figD}. A quite complete experimental data set exists for this
situation as well, both for the HFF (Fig.~\ref{fig10}-a) and the EFG (Fig.~\ref{fig10}-b)
(data are copied from Ref.~\onlinecite{Brewer1990}, the original data are in
Ref.~\onlinecite{Bleaney1972}).
In the experiments, the ions were not really free but were incorporated in a paramagnetic
salt, and the effect of crystal fields was removed later in order to find the free ion
values. By the same procedure as explained for lanthanides in Fe, we extracted LDA+U
predictions for the HFF and $V_{\mathrm{zz}}$ for divalent and trivalent lanthanide ions,
which are given by dashed and full lines in Fig.~\ref{fig10}, respectively. For lanthanides
in Fe, the positive z-direction was naturally defined by the moments of the ferromagnetic
Fe host atoms. For free lanthanides, the total (=spin+orbital) angular momentum $J$
determines the positive z-direction. Due to this different choice of axes, there is an
\emph{apparent} sign change for the heavy lanthanides between Fig.~\ref{fig1} (in Fe) and
Fig.~\ref{fig10}-a (free). The agreement with experiment is again quite nice. Eu is
divalent (it was trivalent in Fe), while Yb is trivalent (it was divalent in Fe). For Sm
there is a large deviation, both for the HFF and $V_{\mathrm{zz}}$. But this is no
surprise: it is well-known there are low-lying excited states in Sm which will mix with the
ground state, such that our procedure which is based on Hund's rules ground states is
expected to fail.

\subsection{Non-collinear Magnetism} The possibility to consider
non-collinear magnetism at every infinitesimal region of space has
recently been implemented\cite{Laskowski2004} in the WIEN2k code.
In principal this can be an important feature even for collinear
antiferromagnets as we are dealing with here: it allows the spin
moment to turn \emph{gradually} from the Fe-orientation to the
opposite lanthanide-orientation, and this is a better replication
of what happens also in nature. We did not attempt a full study,
but calculated HFF and EFG for Tm in Fe only. All technical
parameters were chosen exactly the same as in the collinear
calculations. For Tm in Fe as test example, the total HFF changes
from -822 T in the collinear LDA+U calculation to -838 T in a
non-collinear one, while the EFG remains exactly the same: -38.1
$10^{21}$ V/m$^2$. Such a change of 16~T is not small in absolute
value, but is negligible compared to the large values of the HFF's
in this problem. Therefore we conclude that non-collinear
magnetism doesn't play an important role for lanthanides in Fe.

\section{Conclusions and Outlook}
We have demonstrated that LDA leads to a qualitatively wrong behavior for the magnetism of
lanthanides in Fe. Using LDA+U, qualitative and quite reasonable quantitative agreement
with experiment is obtained, both for HFF and EFG. This shows that the semi \emph{ab
initio} LDA+U method is a useful tool even for such sensitive quantities as hyperfine
parameters of strongly correlated impurities in an itinerant magnetic host. We could come
to these conclusions only after applying a careful strategy in order to cope with the lack
of a good criterion to determine the true ground state if LDA+U is used. For all
lanthanides the 4f spin moment couples antiferromagnetically to the Fe 3d moment, in
agreement with the model of Campbell and Brooks. The orbital HFF is by far the dominant
contribution to the total HFF (Figs.~\ref{fig1} and~\ref{fig4}). The calculated value of
the EFG for lanthanides in Fe agrees well with the few experimental data (Fig.~\ref{fig8}).
We discovered an unexpectedly strong contribution of the completely filled 5p shell to the
dipolar HFF and the EFG, which can be explained by their common $1/r^3$ dependence: small
deformations of the 5p shell in a region close to the nucleus are strongly emphasized. A
reinterpretation of recent EFG calculations\cite{Laskowski2004} for Uranium in UO$_2$
suggests that the same is true for the 6p shell in actinides. Furthermore, we conclude that
Yb is divalent in an Fe host, while all other lanthanides are trivalent (including Eu, with
perhaps an exception for Tm). The lightest lanthanides (at least up to Pr) show delocalized
4f behavior, and we conclude that the delocalization-localization transition that typically
happens already at Ce is postponed for lanthanides in Fe: it falls at least after Pr and
current experiments do not exclude that it could go up to Sm (although Pm is certainly
localized). This can be explained by the large effective pressure that is felt by these
lanthanide impurities (either 25 or 40~GPa - Fig.~\ref{pressure}), leading to a larger
overlap between the 4f wave functions and the neighboring Fe-3d. The question of a
postponed localization transition has never been touched before in the 40 years of
experiments on this system. This illustrates what can be the added value of ab-initio
calculations for hyperfine interactions studies. Also in the case of free lanthanide ions,
HFF and EFG can be quantitatively reproduced. Remarkably, Eu is divalent in this case and
Yb is trivalent -- just the opposite as for lanthanides in Fe. The effect of fully
non-collinear magnetism on this problem was tested to be negligible.

Obviously, the numerical agreement of the calculated hyperfine fields with experiment is
for these lanthanides still much less satisfactorily then it is for lighter impurities.
With the current methods, there are only limited possibilities to improve the accuracy of
the calculations. The supercell can be extended to e.g.\ 32 atoms and relaxation of the
Fe-neighbours can be calculated for every individual element (this requires the calculation
of forces including spin-orbit coupling and LDA+U, which is time-consuming and not yet
fully implemented in WIEN2k). But as inevitably a rather arbitrary choice remains to be
made for the value of $U$, it is not clear whether these sophistications will really
improve the agreement with experiment. And most likely they will not add anything new to
the physical insight. In our opinion, new progress in this topic will have to come from
experiment. Many of the experimentally determined HFF's and EFG's carry still large error
bars. Accurate measurements -- for instance with the NMR/ON method -- are desirable (note
that NMR/ON has not yet been applied for any of the lanthanides with a large EFG and/or
HFF: such large hyperfine interactions put severe requirements on the equipment). The
predicted HFF's and EFG's from this work should allow to reduce considerably the frequency
domain that has to be scanned in an NMR/ON experiment, and warrants a more physical and
reliable interpretation of the observed resonances. As most worthwhile experiments, we
suggest a more accurate determination of HFF and EFG for Pr to Sm: this would allow to
examine experimentally the position of the delocalization-localization transition. Once a
data set with improved accuracy will be available, it can serve in its turn as a testing
ground for future generations of \emph{ab initio} many body methods.

\acknowledgments

Illuminating discussions with Th.~Mazet (Nancy), P.~Blaha
(Vienna), P.~Nov\'{a}k (Prague), M.~Divi\v{s} (Prague), B. Barbara
(Grenoble), N.~Severijns (Leuven) and P.~Schuurmans (SCK/CEN -
Mol, Belgium) are gratefully acknowledged. The calculations were
performed on a pc-cluster in Leuven, in the frame of projects
G.0239.03 of the \textit{Fonds voor Wetenschappelijk Onderzoek -
Vlaanderen} (FWO), the Concerted Action of the KULeuven
(GOA/2004/02) and the Inter-University Attraction Pole (IUAP
P5/1). The authors are indebted to L.~Verwilst and J.~Knuts for
their invaluable technical assistance concerning this pc-cluster.

\end{document}